\documentclass{bioinfo}
\usepackage{natbib}
\usepackage{url}
\newcommand{\C}{\mathcal{C}}

\DeclareMathOperator{\rec}{rec}
\DeclareMathOperator{\prc}{prec}
\DeclareMathOperator{\TP}{TP}

\usepackage{graphicx,subfigure}

\copyrightyear{2012}
\pubyear{2012}

\begin{document}
\firstpage{1}

\title[Context-specific transcriptional regulatory network
inference]{Context-specific transcriptional regulatory network
  inference from global gene expression maps using double two-way
  \textit{t}-tests}

\author[J. Qi and T. Michoel]{Jianlong Qi\,$^{1}$ and Tom
  Michoel\,$^{1,2}$\footnote{to whom correspondence should be
    addressed}} 

\address{$^1$School of Life Sciences -- LifeNet, Freiburg Institute
  for Advanced Studies (FRIAS), University of Freiburg, Albertstrasse
  19, D-79104 Freiburg im Breisgau, Germany, $^2$Division of Genetics
  and Genomics, The Roslin Institute, University of Edinburgh, Easter
  Bush, Midlothian EH25 9RG, UK.}

\history{Received on XXXXX; revised on XXXXX; accepted on XXXXX}

\editor{Associate Editor: XXXXXXX}

\maketitle

\begin{abstract}

  \section{Motivation:} Transcriptional regulatory network inference
  methods have been studied for years. Most of them relie on complex
  mathematical and algorithmic concepts, making them hard to adapt,
  re-implement or integrate with other methods. To address this
  problem, we introduce a novel method based on a minimal statistical
  model for observing transcriptional regulatory interactions in noisy
  expression data, which is conceptually simple, easy to implement and
  integrate in any statistical software environment, and equally well
  performing as existing methods.

  \section{Results:} We developed a method to infer regulatory
  interactions based on a model where transcription factors (TFs) and
  their targets are both differentially expressed in a gene-specific,
  critical sample contrast, as measured by repeated two-way
  $t$-tests. Benchmarking on standard \textit{E. coli} and yeast
  reference datasets showed that this method performs equally well as
  the best existing methods. Analysis of the predicted interactions
  suggested that it works best to infer context-specific TF-target
  interactions which only co-express locally. We confirmed this
  hypothesis on a dataset of more than 1,000 normal human tissue
  samples, where we found that our method predicts highly
  tissue-specific and functionally relevant interactions, whereas a
  global co-expression method only associates general TFs to
  non-specific biological processes.

\section{Availability:} A software tool called TwixTrix is available
from 
http://twixtrix.googlecode.com. 

\section{Supplementary information} Supplementary Material is
available from http://www.roslin.ed.ac.uk/supplementary-data.

\section{Contact:} \href{tom.michoel@roslin.ed.ac.uk}{tom.michoel@roslin.ed.ac.uk}
\end{abstract}

\section{Introduction}

Transcriptional regulatory networks, which emerge from the
combinatorial regulation of the expression of all genes in an organism
by a limited number of transcription factors (TFs), control the
cellular response to internal and external perturbations. At present,
direct experimental mapping of complete transcriptional regulatory
networks remains infeasible, particularly in higher organisms,
especially since the structure of these networks is itself
condition-dependent \citep{harbison2004,luscombe2004}. A lot of
attention has therefore been devoted to computationally reconstruct
transcriptional regulatory networks from compendia of genome-wide gene
expression measurements in diverse conditions, time points, cell types
or genotypic backgrounds
\citep{NirFriedman02062004,zhu2004,BansalMukesh}. However, despite
many years of research, it still remains a question which
computational methods are most suited to tackle this problem.
Moreover, regulatory network inference remains a task firmly in the
hands of specialists, and network inference algorithms are still not
routinely included in standard statistical software packages, unlike
for instance differential expression testing or co-expression
analysis.  At least in part this is due to the fact that most network
reconstruction methods depend on non-trivial mathematical concepts
such as mutual information \citep{16723010, FaithJeremiahj},
differential equations \citep{RichardBonneau}, biophysical models
\citep{bussemaker2007}, Bayesian networks \citep{Segal2003Nature,
  NirFriedman02062004,zhu2004, AnaghaJoshi02152009}, ensemble methods
\citep{AnaghaJoshi02152009} or machine learning
\citep{10.1371/journal.pone.0012776}. While these complex mathematical
models are well justified in theory, current gene expression datasets
have high levels of noise and are lacking in resolution, making these
models prone to over-fitting. Furthermore, the resulting algorithms
are difficult if not impossible to re-implement, as they often depend
on poorly documented parameter choices and heuristic techniques, for
instance to improve convergence rates or avoid local optima.

In order to address these problems, we propose a novel method which is
based on a minimal statistical model. The model assumes that TFs and
their targets are both differentially expressed in a gene-specific
sample contrast, but it makes no assumption on any functional
relationship, be it linear or non-linear, between the gene expression
profiles of TFs and their targets. It should thus be ideally suited to
infer regulatory interactions from noisy, low-resolution gene
expression maps. First, the method identifies for each gene its
\emph{critical contrast}, the separation of samples into two sets
across which that gene is most significantly differentially expressed
(as determined by two-way $t$-tests). Secondly, the method takes a
list of TFs or other regulatory proteins, and calculates their
differential expression in the critical contrast of each possible
target gene (again determined by two-way $t$-tests).  The predicted
network is the list of TF-gene associations, ranked by these $t$-test
$P$-values, either considered as a weighted network or cut off at a
desired significance threshold.

The idea to use $t$-tests to predict regulatory interactions was first
proposed by \cite{qi2011a}. Here, we systematically evaluate the
performance of this double two-way $t$-test procedure using benchmark
expression data \citep{FaithJeremiahj, AudreyP.Gasch12012000} and
networks of known transcriptional regulatory interactions
\citep{SocorroGama-Castro01112008, PedroT.Monteiro01112008} in
\textit{E. coli} and yeast, following standard evaluation protocols
established by the DREAM community \citep{DREAM3}. We found that
double two-way $t$-testing performs as well as the best current
methods, especially in yeast. Next we compared the top-ranked
predictions of each method and found that the $t$-test procedure
identifies a considerably different set of interactions than the other
methods. In particular, whereas the top-ranked predictions of the
other methods tend to exhibit high levels of global co-expression
between the TFs and their predicted targets, interactions found by the
$t$-test procedure tend to only co-express locally and involve TFs
that are only expressed under certain experimental conditions.

We therefore hypothesized that the double two-way $t$-test method is
particularly useful to predict context-specific networks in
multi-cellular organisms. To test this hypothesis we applied it to a
global gene expression compendium containing more than 1,000 samples
from normal human tissues \citep{LukkMargus}.  Although due to a lack
of knowledge in human, a systematic evaluation of the predicted
network is impossible, manual analysis of the top-ranked TFs showed
that the functional enrichment of their predicted targets is indeed
highly consistent with known cell-type specific modes of action for
these TFs.

\begin{methods}
\section{Methods}

\subsection{Critical contrast determination}
\label{sec:crit-contr-determ}

The first step of the double two-way $t$-test procedure consists of
determining the critical contrast for each gene in a gene expression
data matrix. The differential expression of a gene $g$ in a partition
$(\C_1,\C_2)$ of the set of samples in two distinct sets can be
determined by the ordinary $t$-statistic,
\begin{equation}\label{eq:ordinaryt} 
  t=\frac{|\mu_{1}-\mu_{2}|}{\sqrt{\frac{(n_1-1)\sigma_{1}^2+(n_2-1)
        \sigma_{2}^2}{n_1+n_2-2}}   \sqrt{\frac{n_1+n_2}{n_1n_2}}}   
\end{equation}
where $\mu_{1}$ and $\mu_{2}$ are the means of the expression values
of gene $g$ in $\C_1$ and $\C_2$, respectively, and similarly,
$\sigma_{1}$ and $\sigma_{2}$ and $n_1$ and $n_2$ denote the standard
deviations and the numbers of conditions in $\C_1$ and $C_2$,
respectively. The critical contrast of $g$ is defined as the partition
$(\C_1,\C_2)$ with highest value of $t$.
An ordered partition is defined as a partition where all expression
values in $\C_1$ are smaller than all expression values in $\C_2$,
i.e., $\max(\C_1)\leq\min(\C_2)$. For any non-ordered partition, we
can create an ordered one with the same $n_1$ and $n_2$ by repeatedly
swapping $\max(\C_1)$ and $\min(\C_2)$. It is not hard to see that the
$t$-statistic for this ordered partition must be higher than for the
original non-ordered partition. Hence the critical contrast can be
determined by taking the maximum over all ordered partitions, of which
there are only $K-1$ per gene, where $K$ is the total number of
samples in the dataset. A minimal number of samples on each side of
the partition can be set, although the factor
$\sqrt{\frac{n_1+n_2}{n_1n_2}}$ in eq. (\ref{eq:ordinaryt}) ensures
that the critical contrast will automatically be balanced (see
Supplementary Material for details).

\subsection{Scoring of regulatory interactions}
\label{sec:pred-regul-inter}

Given a list of candidate transcription factors or other regulators
and a critical contrast for all possible target genes, we define the
interaction score $t_{f,g}$ between a TF $f$ and target gene $g$ as
the $t$-statistic of $f$ in the critical contrast of $g$. The higher
$t_{f,g}$, the more confident we are about the predicted regulatory
interaction $f\to g$. A confidence $P$-value can be computed from
$t_{f,g}$ using a Student's $t$-distribution with $K-2$ degrees of
freedom, where $K$ is the total number of samples in the dataset.

\subsection{Moderated $t$-statistics and background correction}
\label{sec:use-moderated-t}

Some transformations of the interaction score $t_{f,g}$ are worthwile
to consider. Firstly, because for each gene $g$, the differential
expression of a relatively large number of TFs is tested in the same
critical contrast $(\C_1,\C_2)$, while we expect only few of these TFs
to have signifcantly high differential expression, moderated
$t$-statistics can be used \citep{Gordon2004,limma} to provide a more
stable inference in datasets with a limited number of samples.
Secondly, to make the interaction scores better comparable between
genes with potentially very different critical contrasts, we can apply
a background correction defined as
\begin{equation} \label{eq:1}
  Z_{f,g} = \frac{t_{f,g} - \mu_g}{\sigma_g},
\end{equation}
where $t_{f,g}$ is the ordinary or moderated $t$-statistic interaction
score, and $\mu_g$ and $\sigma_g$ are the mean and standard deviation
of $t_{f,g}$ over all TFs $f$ for a given gene $g$,
respectively. Finally, we can also compute the $t$-statistic of a
target gene $g$ in the critical contrast of TF $f$ and define a
symmetric score $Z^{\text{sym}}_{f,g}=Z_{f,g}+Z_{g,f}$.

\subsection{Algorithm implementation}
\label{sec:algor-impl}

A software tool called TwixTrix is available from our website,
providing two implementations of the double two-way $t$-test
procedure. The first implementation (in R) uses the Limma package
\citep{limma} to calculate moderated $t$-statistics and background
corrected interaction scores and is recommended for datasets with a
small number of samples. The second implementation (in R or Matlab)
encodes the critical contrast determination and interaction scoring
using ordinary $t$-statistics purely as matrix operations. It is
ultra-fast and recommended for datasets with a large number of
samples.

\subsection{Comparison with other network inference methods}
\label{sec:comp-with-other}

We downloaded the latest versions of Inferelator
\citep{RichardBonneau}, CLR \citep{FaithJeremiahj}, LeMoNe
\citep{AnaghaJoshi02152009} and GENIE3
\citep{10.1371/journal.pone.0012776} from their respective homepages
and ran them with default settings. For LeMoNe and Inferelator, which
are module network inference algorithms, we assigned each gene to a
singleton module to obtain a TF-gene regulatory network. As a
baseline, we also reconstructed networks based on the Pearson and
Spearman correlations.  All methods considered provide a ranked list
of predicted regulatory interactions. Keeping the first $k$
interactions, recall and precision are defined as
\begin{align*}
  \rec(k) = \frac{\TP(k)}{N_{\mathrm{ref}}} &&
  \prc(k) = \frac{\TP(k)}{k},
\end{align*}
where $\TP(k)$ is the number of true positives, i.e. the number of
known interactions, among the first $k$ predictions and
$N_{\mathrm{ref}}$ is the total number of known interactions. The area
under the recall-precision curve (AUC) provides a measure for the
performance of each method, see \citep{DREAM3} for details.

\subsection{Gene expression data and reference regulatory networks}
\label{sec:gene-expression-data}

We tested our method on datasets for \textit{E. coli}, yeast and
human. The \textit{E. coli} dataset \citep{FaithJeremiahj} contains
expression values for 4345 genes under 189 conditions. We considered
the same 316 candidate regulators and 1882 differentially expressed
genes (sd$>$0.5) as \cite{Michoel19422680}. Results were evaluated
using the reference network of RegulonDB
\citep{SocorroGama-Castro01112008}. The yeast stress dataset measures
the budding yeast's response to a panel of diverse environmental
stresses \citep{AudreyP.Gasch12012000}. We used the same list of 321
candidate regulators and 2355 differentially expressed genes as
\cite{Segal2003Nature}. Results were evaluated using the reference
network of YEASTRACT \citep{PedroT.Monteiro01112008}. The human
dataset dataset consists of 5,372 samples \citep{LukkMargus}, from
which we selected 1033 samples measuring gene expression in 67 diverse
tissues under normal conditions. We reconstructed networks using a
list of 941 candidate transcription factors from TcoF
\citep{Schaefer21102010} and 12,568 differentially expressed genes
(sd$>$0.5).

\end{methods}

\section{Results and Discussion}

\subsection{Benchmarking on \textit{E. coli} and yeast datasets}

\begin{figure}
  \centering
  \includegraphics[width=\linewidth]{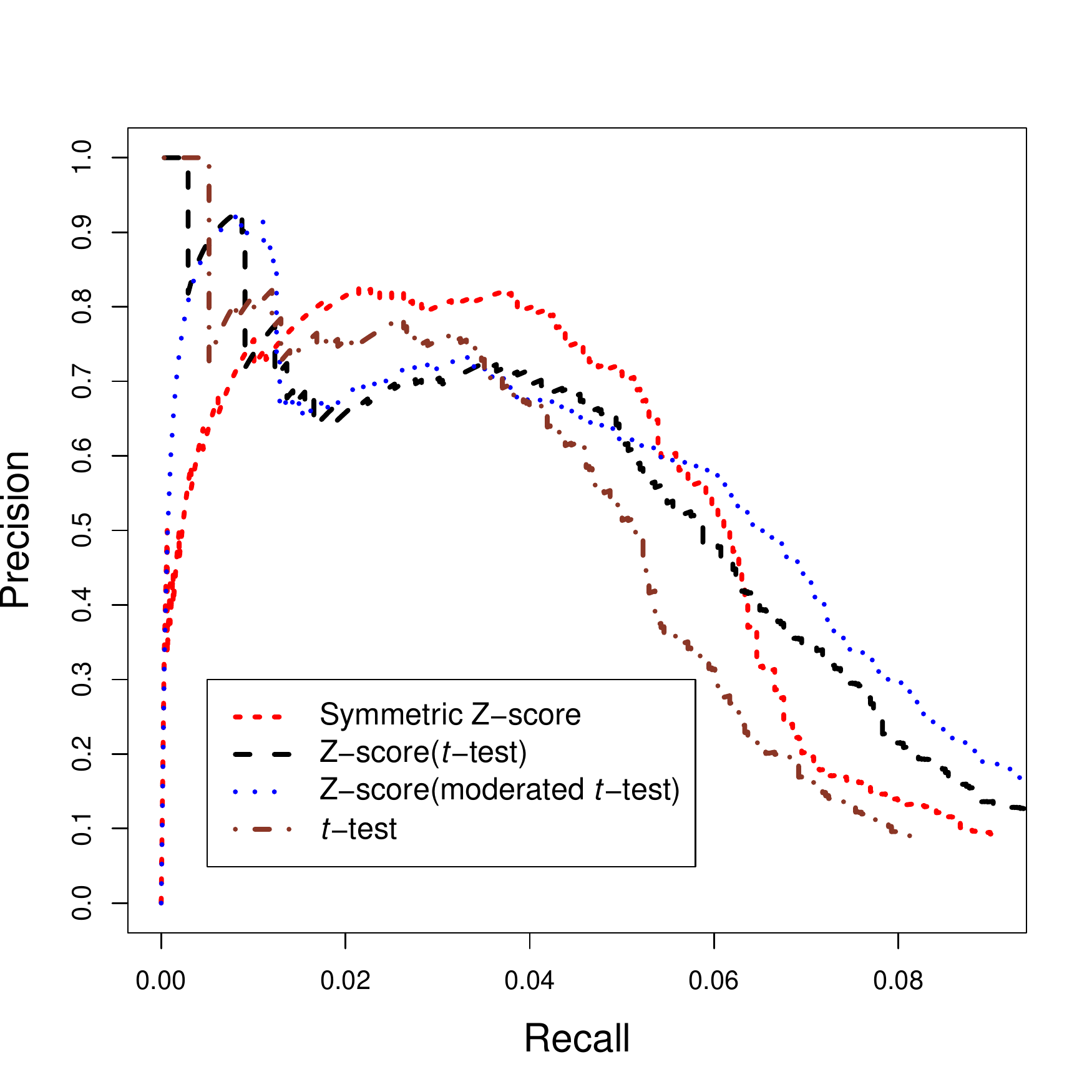}
  \caption{Recall-precision curves for different versions of double
    two-way $t$-test interaction scores in \textit{E. coli}.}
  \label{fig:ttest-comp-ecoli}
\end{figure}

\begin{figure}
  \centering
  \includegraphics[width=\linewidth]{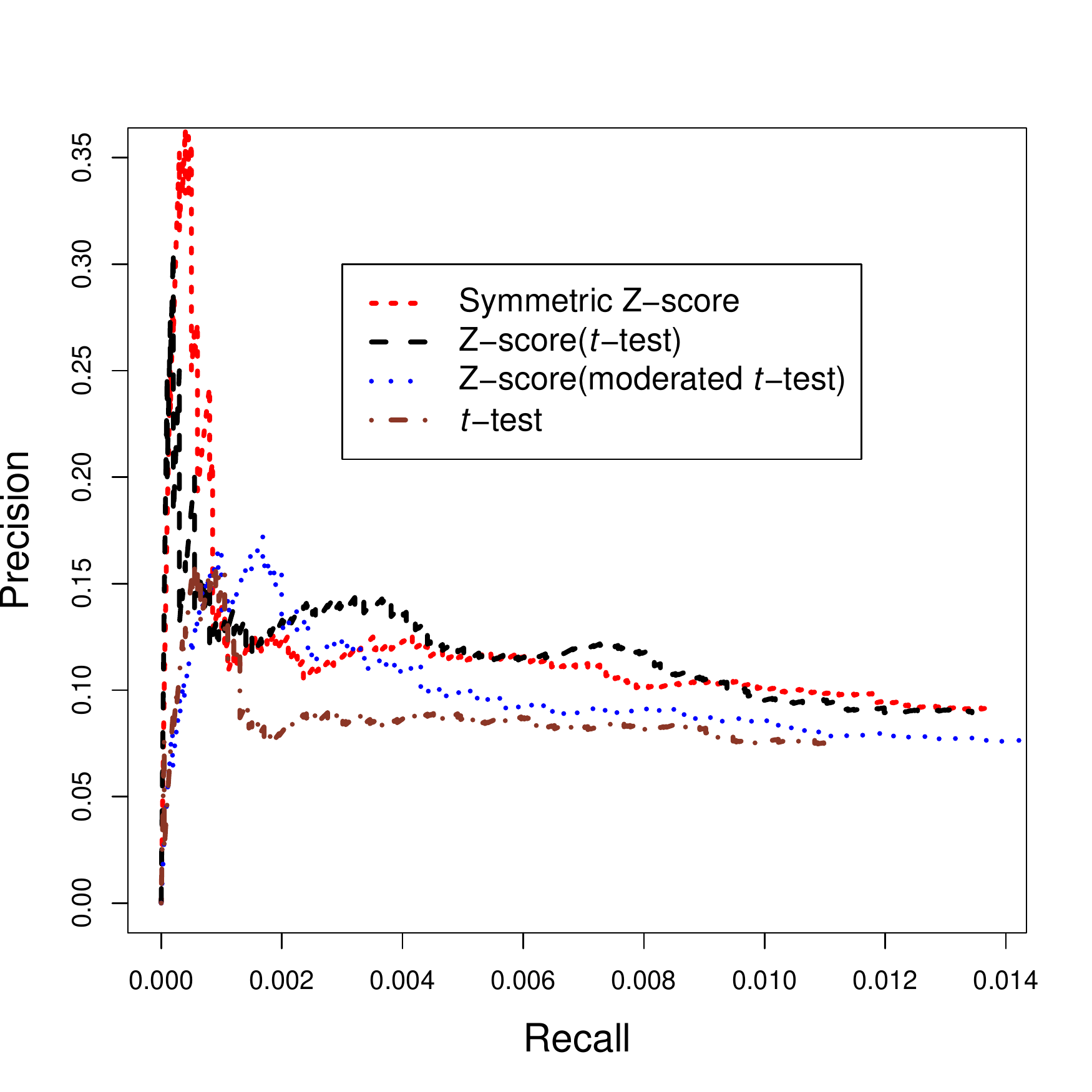}
  \caption{Recall-precision curves for different versions of double
    two-way $t$-test interaction scores in yeast.}
  \label{fig:ttest-comp-yeast}
\end{figure}

\begin{figure}
  \includegraphics[width=\linewidth]{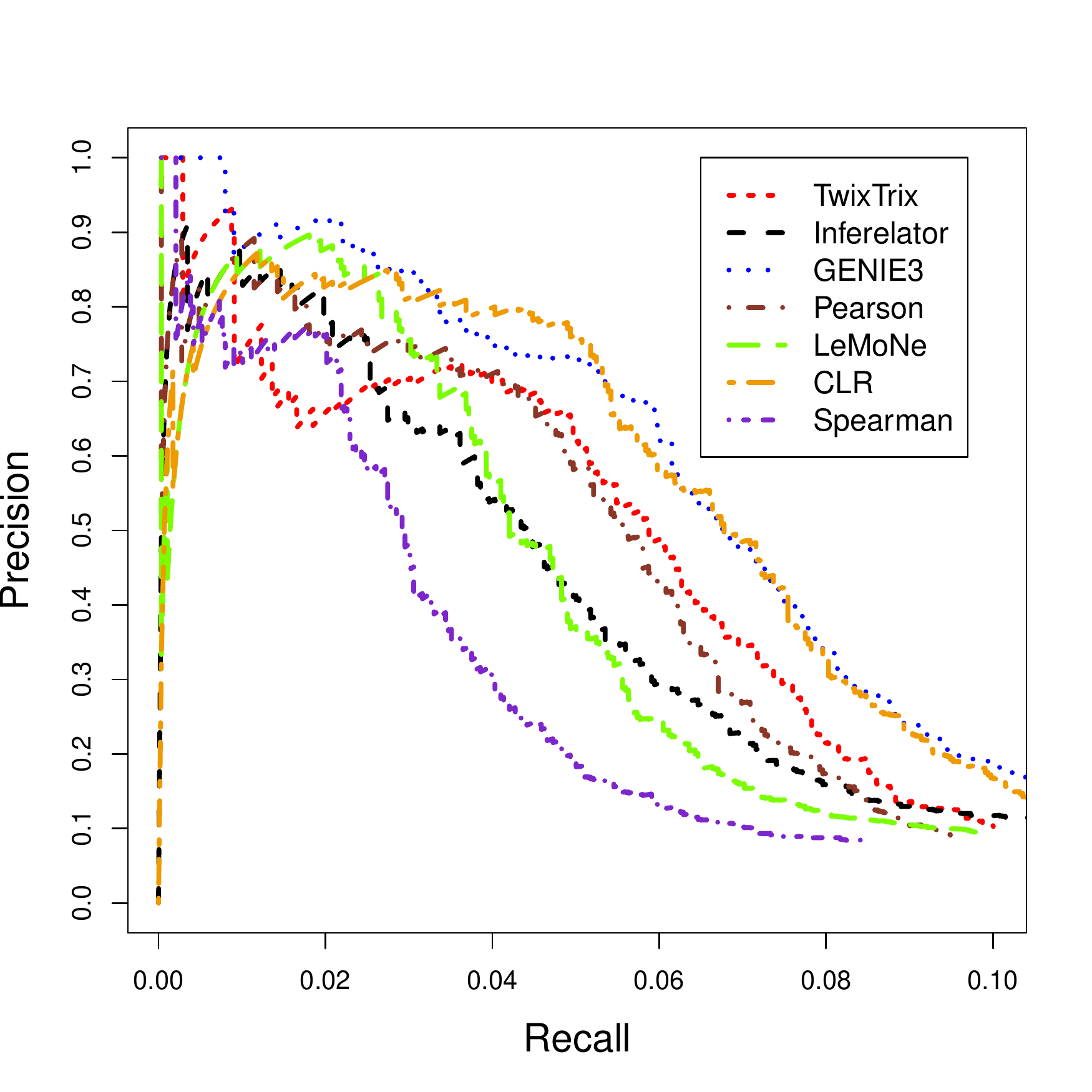}
  \caption{Recall-precision curves for seven transcriptional
    regulatory network inference algorithms in $E. coli$.}
  \label{fig:ecoliprecision}
\end{figure}

\begin{figure}
  \includegraphics[width=\linewidth]{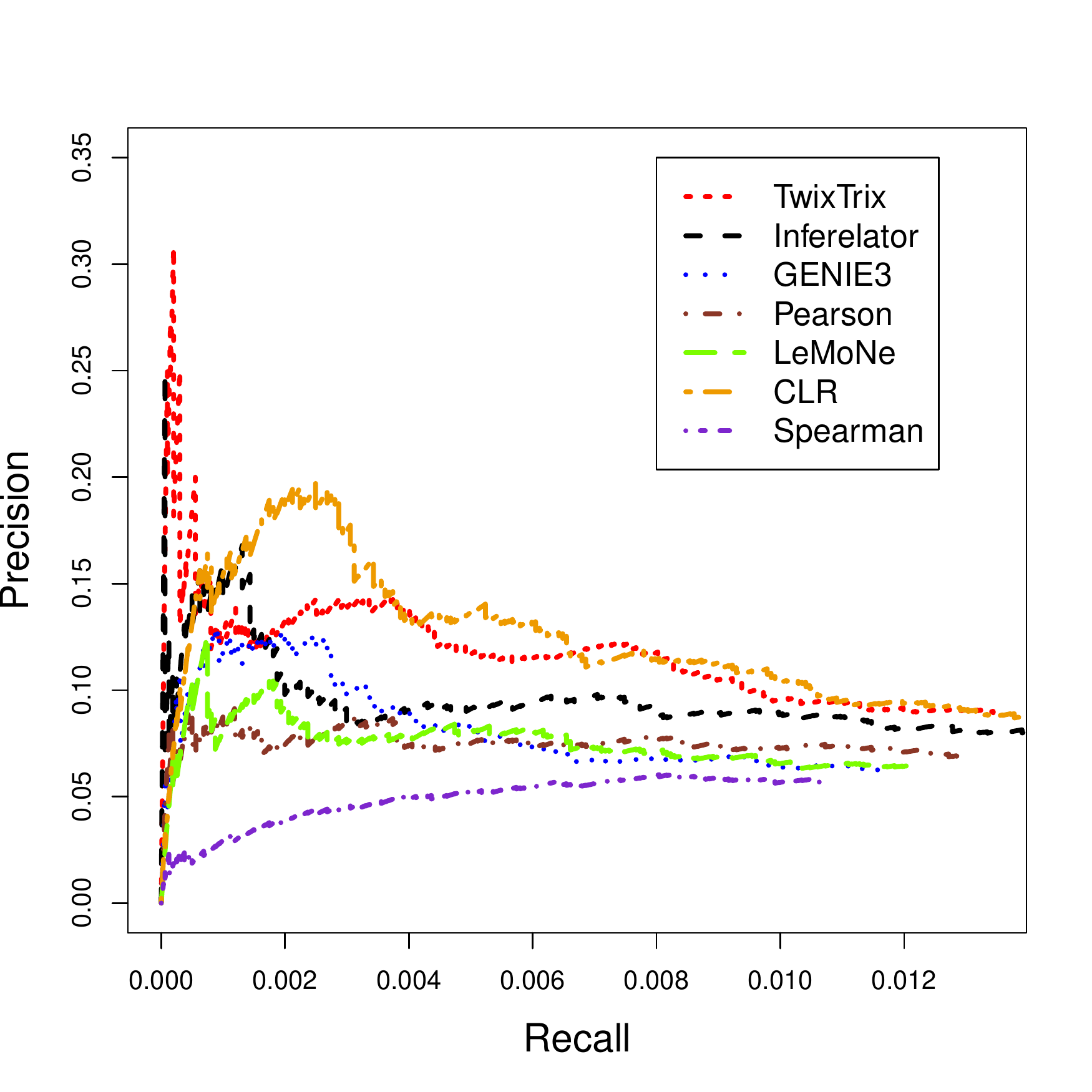}
  \caption{Recall-precision curves for seven transcriptional
    regulatory network inference algorithms in yeast.}
  \label{fig:yeastprecision}
\end{figure}

To benchmark the double two-way $t$-test procedure, we analysed its
performance on standard \textit{E. coli} and yeast datasets, by
calculating recall vs. precision curves based on the top 3000
predicted interactions (see Methods for details). In Figure
\ref{fig:ttest-comp-ecoli} and \ref{fig:ttest-comp-yeast}, we compared
different versions of double two-way $t$-test interaction scores,
namely the $t$-statistic $t_{f,g}$ of a TF $f$ in the critical
contrast of a gene $g$, the background corrected score $Z_{f,g}$ (eq. (\ref{eq:1})) for the ordinary
$t$-statistic and one for the moderated $t$-statistic, and a
symmetrized background corrected score for the ordinary $t$-statistic
(cf. Methods section \ref{sec:use-moderated-t}). Although the
symmetrized version works slightly better in yeast, this is not the
case in \textit{E. coli}. In order not to overfit for a specific
dataset, we choose the unsymmetric background corrected score for the
ordinary $t$-statistic (cf. eq. (\ref{eq:1})) as the default score,
because it performs well in both datasets and is conceptually the
simplest and fastest to compute. All results reported in the
remainder of this paper refer to this interaction score.

Next we compared the double two-way $t$-test procedure against six
other methods on the \textit{E. coli} and yeast datasets, again
calculating recall vs. precision curves based on the top 3000
predicted interactions by each method (Figure \ref{fig:ecoliprecision}
and \ref{fig:yeastprecision}). As has been observed before
\citep{Michoel19422680}, overall performance in yeast compared to
\textit{E. coli} is lower for all methods. This may be due to more
complex regulatory mechanisms in eukaryotes vs. prokaryotes, a less
accurate reference network against which performance is measured or,
most likely, a combination of these two. Most importantly, neither of
the algorithms is better than all the others in both organisms
(cf. Table \ref{table:auc}). TwixTrix, our implementation of the
double two-way $t$-test algorithm, performs equally good as the much
more complicated algorithms, especially in yeast where it is ranked
second best.  Also noteworthy is the fact that in \textit{E. coli},
but not in yeast, regulatory interaction prediction based on the
Pearson correlation between TFs and putative target genes also
performs similarly well as the other methods. This suggests that,
generally speaking, TFs and their targets tend to be globally
co-expressed in prokaryotes and the real challenge is to predict
regulatory networks in eukaryotes.

\begin{table}
  \centering
  \begin{tabular}{lcc}
     & \textit{E. coli} & Yeast\\
    \hline
    TwixTrix & 0.05182 & 0.00157 \\
    Inferelator & 0.04624 & 0.00140 \\
    GENIE3 & \textbf{0.06767} & 0.00097 \\
    LeMoNe & 0.04415 & 0.00091 \\
    CLR & 0.06269 & \textbf{0.00190} \\
    Pearson & 0.05003 & 0.00097 \\
    Spearman & 0.03157 & 0.00052\\
    \hline\\
  \end{tabular}
  \caption{Area under the recall-precision curve for each method in
    \textit{E. coli} and yeast. The bold numbers indicate the highest
    value in each organism.}
  \label{table:auc}
\end{table}

\subsection{TwixTrix identifies context-specific interactions}
\label{sec:twixtr-ident-cont}

An important recent insight has been that different network inference
strategies identify different aspects of a regulatory
system. Understanding how a method differs from others has therefore
become more important than simple recall-precision measurements, with
the eventual goal to build meta-networks which integrate predictions
from diverse computational methods
\citep{Michoel19422680,Marbach06042010}. Here we focused on
characterizing TwixTrix-predicted interactions in yeast, where it is
most successful relative to the other methods. The corresponding
figures for \textit{E. coli} can be found in the Supplementary
Material.

\begin{figure}
  \centering
  \includegraphics[width=\linewidth]{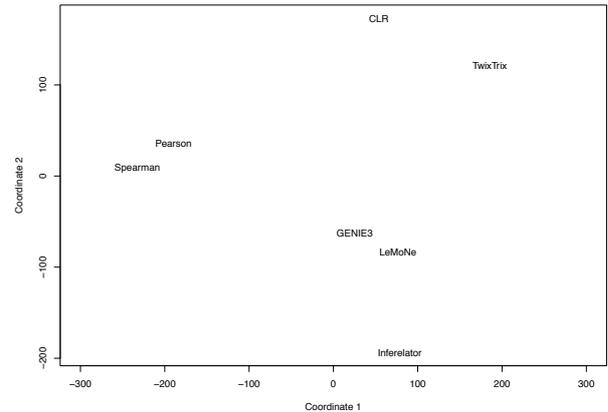}  
  \caption{Multi-dimensional scaling plot, using the number of
    non-overlapping interactions among the top 500 predicted
    interactions as a distance measure between network inference
    methods.}
  \label{fig:mds}
\end{figure}

First we compared the overall similarity of interactions predicted by
each method. The overlap (measured as the number of common
interactions among the top 500 predicted interactions) ranges from 31
to 232 common interactions. TwixTrix shares between 31 (with Spearman
correlation) and 144 (with LeMoNe) interactions with the other
methods. Using the number of non-overlapping interactions as a
distance measure, the relative similarities between each of the seven
network inference methods is visualised in Figure \ref{fig:mds}. As
expected, the networks based on Pearson and Spearman correlation are
most similar. GENIE3 \citep{10.1371/journal.pone.0012776} and LeMoNe
\citep{AnaghaJoshi02152009} both use decision trees to assign
regulators to target genes and consequently predict similar networks
as well. TwixTrix, CLR and Inferelator each occupy a more unique
position in the network inference landscape.

\begin{figure}
  \centering
  \includegraphics[width=\linewidth]{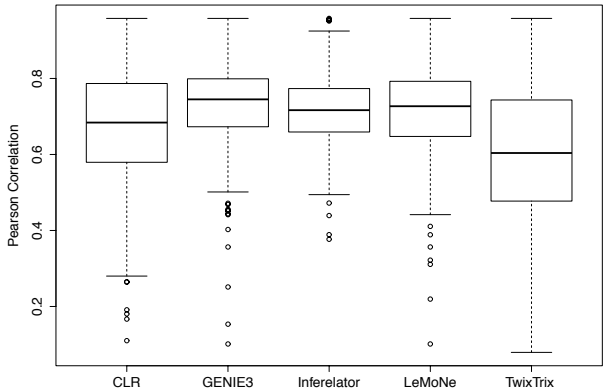}
  \caption{Distribution of Pearson correlations between the top 500
    predicted TF-target interactions in yeast from five network
    inference methods.}
  \label{fig:pc-box}
\end{figure}

Since TwixTrix is based on differential expression testing, we
hypothesized that it tends to identify TF-target interactions which do
not necessarily co-express under all conditions.  Figure
\ref{fig:pc-box} shows the distribution of Pearson correlations
between TFs and their predicted targets for the top 500 interactions
in yeast for each method. Interactions predicted by TwixTrix indeed
have significantly lower Pearson correlations than interactions
predicted by the other methods. GENIE3, Inferelator and LeMoNe, which
perform less well in yeast than TwixTrix and CLR, are especially
biased towards inferring interactions which co-express under most
conditions in the dataset.

\begin{figure*}
  \centering
  \includegraphics[width=\linewidth]{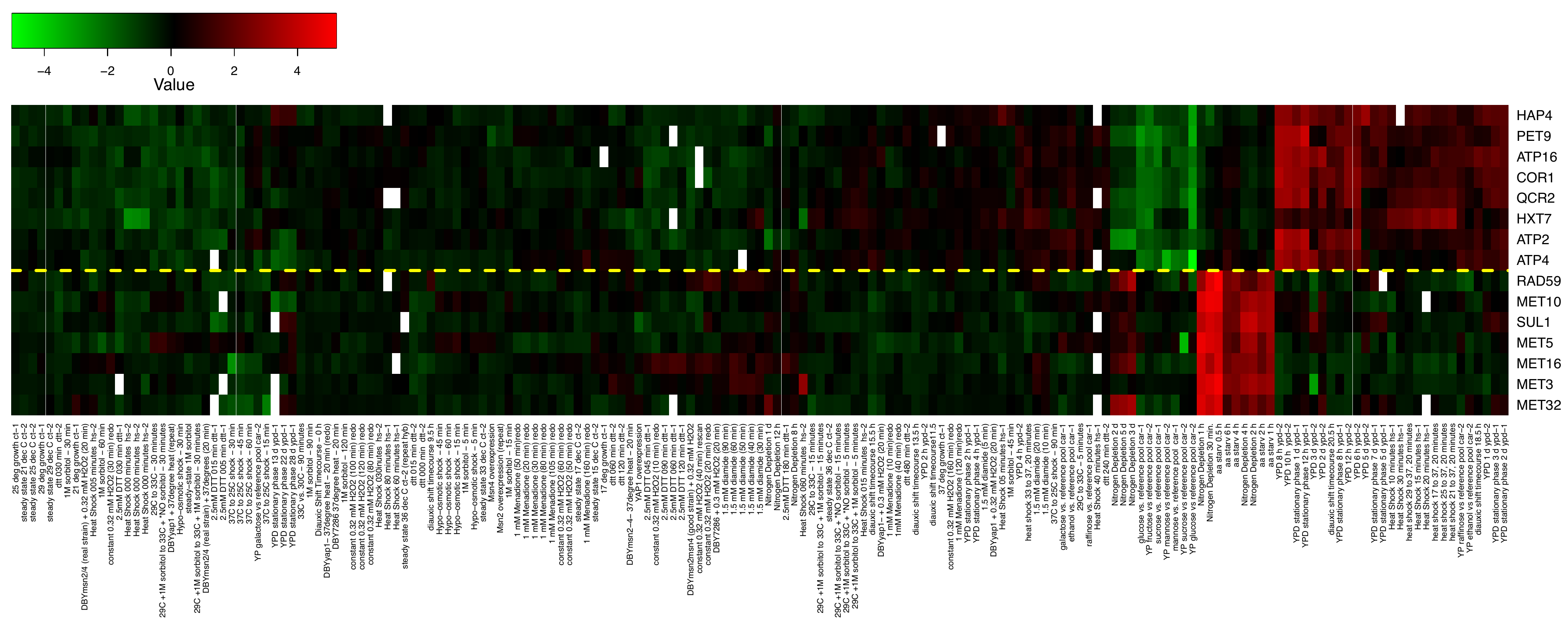}
  \caption{Heatmap showing the expression values of MET32 and HAP4 and
    their respective highest-scoring known targets in the TwixTrix
    network. Red - over-expressed, green - under-expressed and black -
    no change compared to wild-type expression levels.}
  \label{fig:yeast-heatmap}
\end{figure*}

A simple example illustrates the difference between context-specific
and global interactions. MET32 is a transcription factor involved in
the regulation of methionine (an amino acid) biosynthesis. It has 30
predicted targets in the top 500 network of TwixTrix, of which 9 are
known targets, which are strongly enriched for amino acid biosynthesis
genes (hypergeometric $P<10^{-12}$ after multiple testing
correction). The interactions predicted for MET32 are among the
highest scoring by the double two-way $t$-test, yet their global
Pearson correlation is only 0.64 on average. As shown in Figure
\ref{fig:yeast-heatmap}, MET32 and its targets are only expressed
under amino acid starvation and nitrogen depletion, resulting in a
strong signal upon differential expression testing despite weak global
correlation. The opposite situation happens for HAP4, a
transcriptional activator and global regulator of respiratory gene
expression. HAP4 has 14 predicted targets in the top 500 network of
TwixTrix, of which 11 are known targets, which are strongly enriched
for cell death and oxidative phosphorylation (hypergeometric
$P<10^{-10}$ after multiple testing correction). Despite a higher
global co-expression between HAP4 and its predicted targets (average
Pearson correlation 0.73), the $t$-test scores are relatively low
(only two predicted interactions in the top 100). In contrast, for
GENIE3, Inferelator and LeMoNe, which all favor highly co-expressed
interactions (cf. Figure \ref{fig:pc-box}), the HAP4 predictions are
all among the highest scoring true positive interactions (respectively
4, 7 and 4 true positives in the top 10 predictions). Figure
\ref{fig:yeast-heatmap} shows that HAP4 and its targets are all highly
expressed under YPD stationary phase and heat shock conditions, while
they are under expressed in glucose conditions. This results in a
strong global, but weaker condition-specific signal.

\subsection{Tissue-specific network inference from a human gene
  expression atlas}

While the ability to detect context-specific interactions in yeast is
important, the dataset consisted of samples for altogether ten
different experimental conditions \citep{AudreyP.Gasch12012000},
making the distinction between condition-specific and global
interactions somewhat arbitrary. In contrast, global gene expression
maps in mammalian systems can consist of hundreds of different cell
and tissue types, developmental stages, or disease states
\citep{LukkMargus}. To test TwixTrix in such a setting, we applied it
to a large dataset of more than 1,000 samples from 67 normal human
tissues \citep[see Methods for details]{LukkMargus}. 

Because there exists no comprehensive reference database against which
an inferred transcriptional regulatory network can be validated in
human, we manually analysed the top-ranking transcription factors and
computed the functional enrichment of their targets in the top 10,000
TwixTrix-interactions (Table \ref{table:modules}). In all cases, the
TFs are expressed only in a very small set of samples from specific
tissues, which are highly consistent with the most enriched GO term
among their targets. This supports the hypothesis that double two-way
$t$-testing is indeed well-suited to predict tissue-specific gene
regulatory networks.

\begin{table*}
  {\tiny
    \begin{center}
      \begin{tabular}{|l|p{6cm}|p{3cm}|p{0.3cm}|p{0.3cm}|p{0.3cm}|p{3cm}|p{0.8cm}|}
        \hline
        Regulator & Description of Regulator & Context & $N$ & $M$ &
        $m$ & Enriched GO Term & $P$-value \\  
        \hline
        GCM1 & The placental transcription factor glial cell missing 1
        (GCM1) regulates expression of syncytin-1 and -2 fusogenic
        proteins \citep{chang20113820}. & Placenta basal plate samples
        & 73 & 64 & 20 & Female pregnancy & 5.94E-22  \\ 
        \hline
        KLF1 & This gene encodes a hematopoietic-specific
        transcription factor that induces high-level expression of
        adult beta-globin and other erythroid genes
        \citep{Perseu20102011}. & Blood, bone, and fetal blood samples
        & 21  & 20 & 3 & Regulation of symbiosis, encompassing
        mutualism through parasitism & 2.36E-4 \\ 
        \hline
        MYT1L & Myelin transcription factor 1-like (MYT1L) is a member
        of the myelin transcription factor 1 and plays a role in the
        development of the nervous system \citep{wang201013622}. &
        Brain related samples & 482 & 421 & 95 &  Transmission of
        nerve impulse & 8.28E-42\\ 
        \hline
        MYF6 & The protein encoded by this gene is a probable basic
        helix-loop-helix (bHLH) DNA binding protein involved in muscle
        differentiation \citep{braun1990821}. & Bone, quadriceps
        muscle, skeletal muscle, and vastus lateralis samples & 45 &
        42 & 24 & Muscle system process & 6.14E-30\\ 
        \hline
        NR1I2 & NR1I2 plays a central role in regulating liver and
        gastrointestinal drug metabolism \citep{wang20103220}. & Small
        intestine samples & 44 & 40 & 6 & Digestion & 7.43E-5 \\ 
        \hline
        NR5A2 & NR5A2 is expressed primarily in liver, intestine, and
        pancreas, where it regulates expression of proteins
        maintaining cholesterol homeostasis \citep{Benod11102011}. &
        Kidney samples  & 41 & 37 & 5& Drug metabolic process & 3.17E-6 \\ 
        \hline 
        PAX2 & Mutations within PAX2 have been shown to result in
        optic nerve colobomas and renal hypoplasia
        \citep{Karafin20111264}. & Kidney samples & 68 & 67 & 19 &
        Organic acid metabolic process & 2.50E-6\\ 
        \hline
        PAX9 & PAX9  may involve development of stratified squamous
        epithelia as well as various organs and skeletal elements
        (\citeauthor{ncbi}). & Esophagus, hypopharynx, oropharynx, and tonsil
        samples & 52 & 46 & 16 & Epidermis development & 5.71E-16 \\ 
        \hline
        PITX1 & PITX1 may involve development of oral cancer
        \citep{Tatiana2011225}. & Esophagus,  hypopharynx, oropharynx,
        and tonsil samples & 76 & 69 & 16 & Epidermis development &
        5.71E-16 \\ 
        \hline
        TBX5 & TBX5 may play a role in heart development and
        specification of limb identity \citep{Sotoodehnia20111061}. &
        Atrial myocardium and cardiac ventricle samples & 91 & 87 & 29
        & Muscle contraction & 9.75E-29 \\ 
        \hline
        NANOG & Nanog is a core factor that is required for the
        maintenance of embryonic stem (ES) cell pluripotency and
        self-renewal \citep{Das09122011}. & Human universal reference
        and kidney samples & 41 & 34 & 12 & Response to wounding &
        5.15E-4 \\  
        \hline
      \end{tabular}
    \end{center}
  }
  \caption{Transcription factors with highest-scoring interactions in
    the TwixTrix-predicted regulatory network in human. Context -
    samples in critical contrast; $N$ - number of targets in top 10,000
    predictions; $M$ - number of targets with GO annotation; $m$ -
    number of targets annotated with most significant GO term; Enriched
    GO term - most significantly enriched GO term; $P$-value -
    hypergeometric enrichment $P$-value after correction for multiple testing.}
\label{table:modules}
\end{table*}

\begin{figure}
  \centering
  \includegraphics[width=\linewidth]{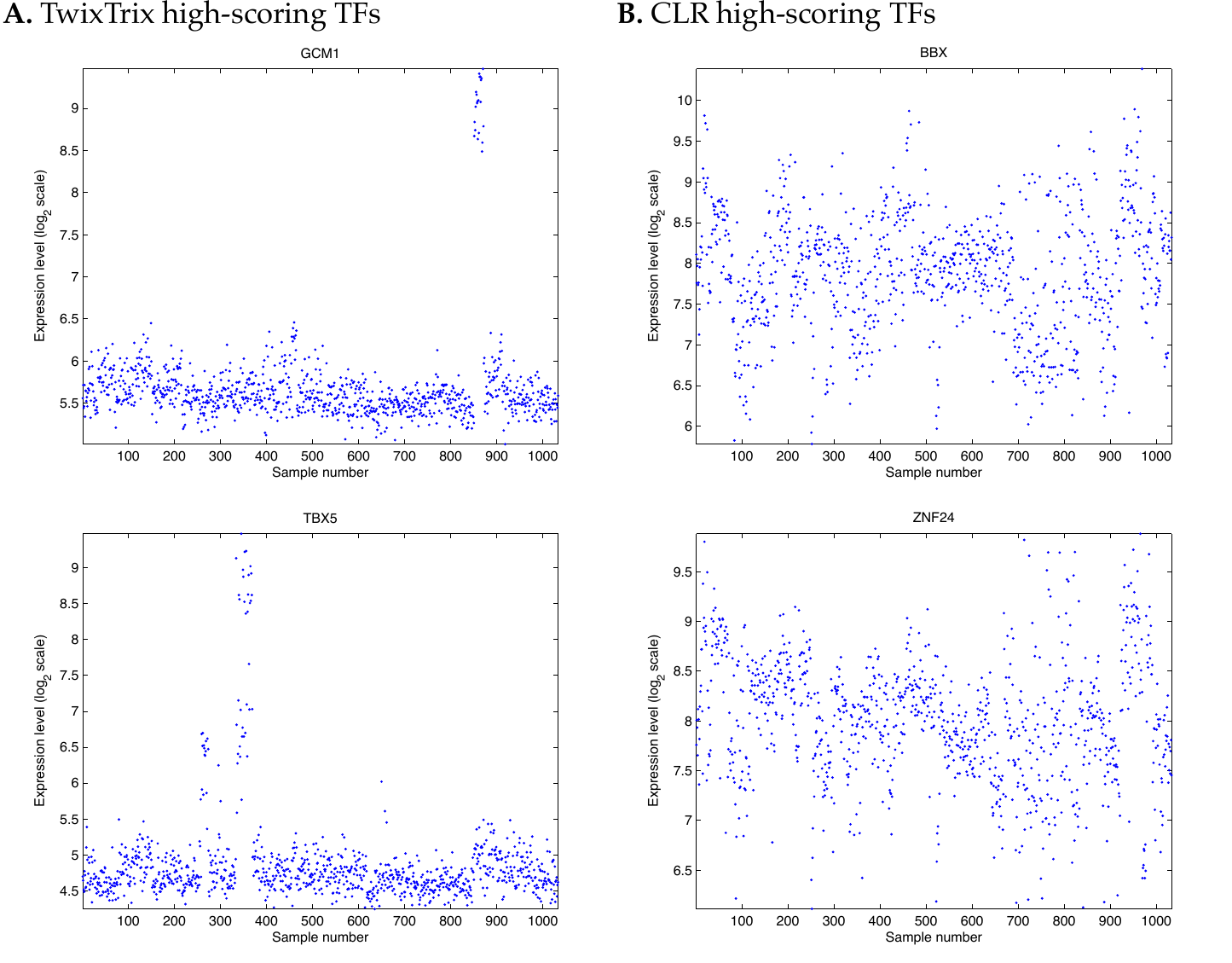}
  \caption{Absolute $\log_2$ expression profiles across 1,033 normal
    human tissue samples for \textbf{(A)} two high-scoring TFs in the
    TwixTrix network and \textbf{(B)} two high-scoring TFs in the CLR
    network.}
  \label{fig:human-tfs}
\end{figure}

\begin{figure}
  \centering
  \includegraphics[width=\linewidth]{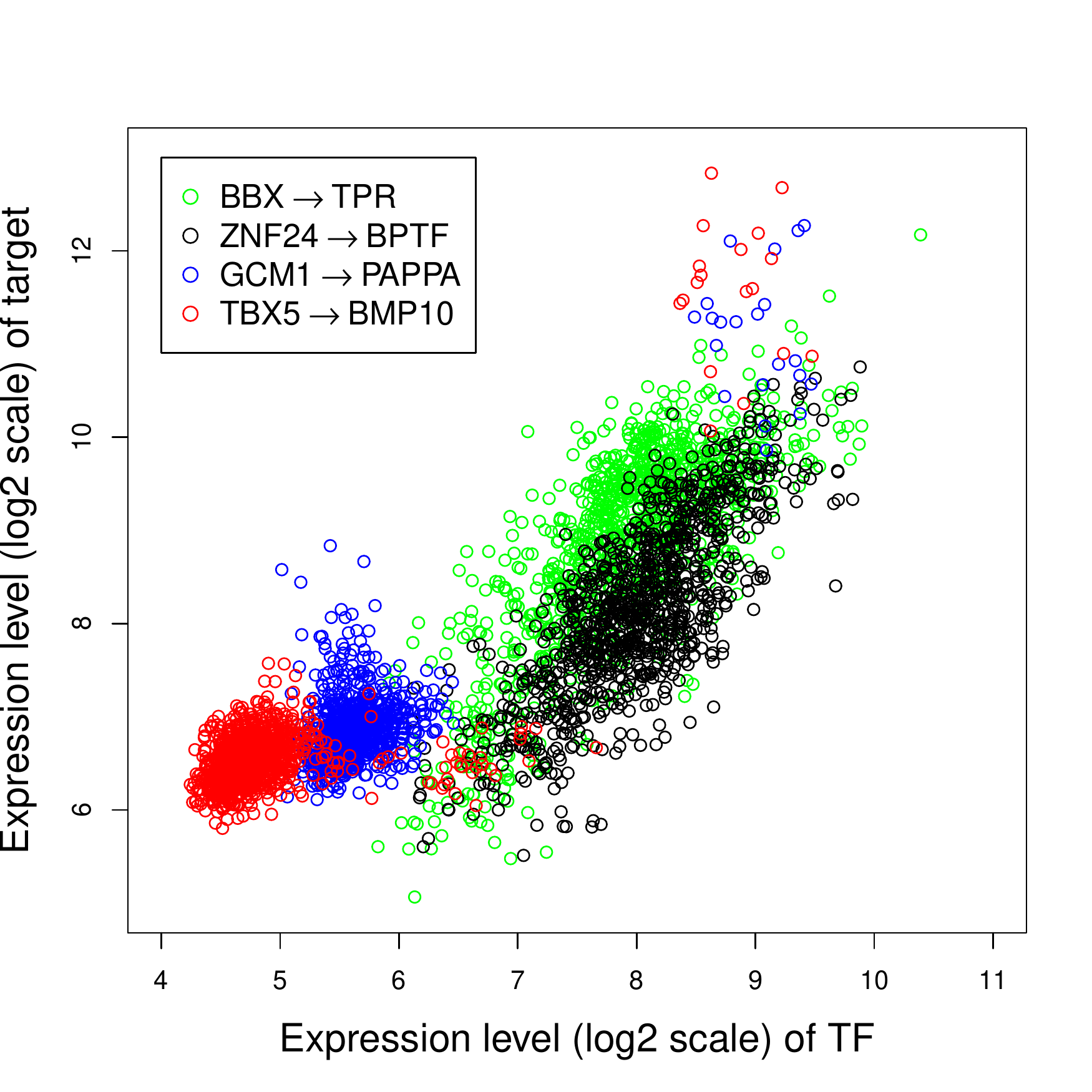}
  \caption{Scatter plot of absolute $\log_2$ expression levels for
    representative high-scoring TwixTrix (blue and red) and
    high-scoring CLR (green and black) predicted interactions.}
  \label{fig:human_tf_target}
\end{figure}

As expected, the different characteristics observed in yeast between
TwixTrix and the other, more globally oriented methods become much
more pronounced in the human dataset. Figure \ref{fig:human-tfs}A
shows the expression profiles across all samples for two
representative TFs from Table \ref{table:modules}. GCM1 is a TF
necessary for placental development. It and its predicted targets (see
high-resolution heatmap in Supplementary Material) are highly
expressed in placental samples (Figure \ref{fig:human-tfs}A,
top). Likewise, TBX5, a TF with a role in heart development, and its
predicted targets (see high-resolution heatmap in Supplementary
Material) are highly expressed in samples from the heart (Figure
\ref{fig:human-tfs}A, bottom). For both TFs, we took a representative
high-scoring target and created a scatter plot of their respective
expression levels (Figure \ref{fig:human_tf_target}, blue and red
points). These TF-target relations have a well-defined tissue-specific
off/on behavior. These highly tissue-specific co-expression signals
result in very significant $t$-test scores. However, because of the
noisy low-level expression in all the other samples, such a signal
cannot be detected by global co-expression methods. In contrast,
Figure \ref{fig:human-tfs}B shows the expression profiles across all
samples for two representative high-scoring TFs in the CLR
network. Both have a characteristic profile which fluctuates across
different tissue types. BBX (Figure \ref{fig:human-tfs}B, top) is a TF
necessary for cell cycle progression from G1 to S phase and ZNF24
(Figure \ref{fig:human-tfs}B, bottom) is a TF involved in promoting
the cell cycle in the developing central nervous system. For both of
them, their CLR-predicted target sets (see high-resolution heatmaps in
Supplementary Material), though highly co-expressed across most
samples, are only enriched for non-specific functional categories such
as regulation of gene expression or metabolic process. A scatter plot
of TF-target expression levels for a representative target for both
TFs confirms that they show a high linear correlation across most
samples (Figure \ref{fig:human_tf_target}, green and black
points). CLR predictions thus clearly represent general processes
which are globally co-expressed and not confined to a single tissue or
cell type.

GENIE3, Inferelator and LeMoNe could not be applied with reasonable
runtime on the complete human dataset. We therefore reduced the size
of the dataset by averaging samples from the same tissue type. Results
on this reduced dataset confirmed that all methods except TwixTrix
give highest rank to globally co-expressed TF-target pairs involved in
general cellular processes, although the relation between the
expression of TFs and their predicted targets tends to be more
non-linear for GENIE3 and LeMoNe than for CLR and Inferelator (see
Supplementary Material for details, including a runtime comparison
between all algorithms). As a particular example of
non-tissue-specific interactions, GENIE3 and LeMoNe predict 35 and 36
targets, respectively, among their top 200 predictions for the TF
FOXM1, which are strongly enriched for M phase ($P<10^{-14}$) and
mitotic cell cycle ($P<10^{-15}$), respectively. FOXM1 is a
transcriptional activator involved in cell proliferation which is
indeed known to regulate the expression of several cell cycle genes.

\section{Conclusion}

Reconstructing transcriptional regulatory networks from genome-wide
gene expression data remains an important bioinformatics
challenge. Although diverse mathematical and computational methods
have been proposed to address this problem, they have not been as
successful as might originally have been expected. A possible reason
is that current gene expression datasets are too noisy and lack the
resolution for adequately fitting complex mathematical models.  Here
we analysed a method which, rather than adding to the complexity of
network inference methods, uses a minimal statistical model for
associating transcription factors to putative target genes without
assuming any linear or non-linear functional relationship between
their expression profiles. The method is based on a double two-way
$t$-test which assesses the differential expression of a TF in the
critical sample contrast of all genes. Essentially, this results in a
local co-expression measure which appears well-suited to capture
context-specific transcriptional activity, at the expense of giving
less weight to globally co-expressed TF-target pairs.

In bacteria, much of the cellular response to perturbations is
controlled at transcriptional level only, such that many TF-target
pairs are co-expressed under all experimental conditions. Here the
double two-way $t$-test therefore does not improve upon existing
methods. In yeast however, there is evidence of known transcriptional
interactions which only co-express under specific conditions. The
$t$-test procedure prioritizes such interactions and indeed performs
better in yeast than \textit{E. coli}, relative to the existing methods.

Taking this one step further, we hypothesize that the double two-way
$t$-test method for inferring regulatory interactions will be
particularly useful to analyse global gene expression maps in
multi-cellular organisms which combine data from hundreds of different
samples. Indeed we confirmed that our method predicts highly
tissue-specific and functionally relevant interactions from a dataset
of more than 1,000 normal human tissue samples, whereas global
co-expression methods only associate general TFs to non-specific
biological processes.

In view of the time it takes to experimentally generate large
expression compendia, judging a network inference method by its
running time is perhaps not very relevant. Nevertheless we note that,
depending on hardware details, the $t$-test procedure took not more
than a few seconds to analyse the human dataset, while the other
methods needed from a few hours upto several days. Having a fast
method is clearly beneficial, e.g. to easily compare results from
different normalizations of the data, or from different subsets of a
large data compendium, e.g. from normal vs. disease states, cell lines
vs. tissue samples, etc.

In summary, we believe that the double two-way $t$-test method
provides a useful addition to existing network inference methods,
whose primary strength lies in prioritizing context-specific
regulatory interactions from global gene expression maps which
integrate data from hundreds to thousands of samples from diverse
experimental treatments, cell types, tissues, developmental stages or
individuals.

\bibliographystyle{natbib}

\begin{thebibliography}{}

\bibitem[Bansal {\em et~al.}(2007)Bansal, Belcastro, Ambesi-Impiombato, and
  di~Bernardo]{BansalMukesh}
Bansal, M., Belcastro, V., Ambesi-Impiombato, A., and di~Bernardo, D. (2007).
\newblock {How to infer gene networks from expression profiles}.
\newblock {\em Molecular Systems Biology\/}, {\bf 3}, 78--87.

\bibitem[Benod {\em et~al.}(2011)Benod, Vinogradova, Jouravel, Kim, Fletterick,
  and Sablin]{Benod11102011}
Benod, C., Vinogradova, M.~V., Jouravel, N., Kim, G.~E., Fletterick, R.~J., and
  Sablin, E.~P. (2011).
\newblock Nuclear receptor liver receptor homologue 1 (lrh-1) regulates
  pancreatic cancer cell growth and proliferation.
\newblock {\em Proceedings of the National Academy of Sciences\/}, {\bf
  108}(41), 16927--16931.

\bibitem[Bonneau {\em et~al.}(2006)Bonneau, Reiss, Shannon, Facciotti, Hood,
  Baliga, and Thorsson]{RichardBonneau}
Bonneau, R., Reiss, D.~J., Shannon, P., Facciotti, M., Hood, L., Baliga, N.~S.,
  and Thorsson, V. (2006).
\newblock {The Inferelator}: an algorithm for learning parsimonious regulatory
  networks from systems-biology data sets \textit{de novo}.
\newblock {\em Genome Biology\/}, {\bf 7}(5), R36.

\bibitem[Braun {\em et~al.}(1990)Braun, Bober, Winter, Rosenthal, and
  Arnold]{braun1990821}
Braun, T., Bober, E., Winter, B., Rosenthal, N., and Arnold, H.~H. (1990).
\newblock {Myf-6}, a new member of the human gene family of myogenic
  determination factors: evidence for a gene cluster on chromosome 12.
\newblock {\em The EMBO Journal\/}, {\bf 9}(3), 821 -- 831.

\bibitem[Bussemaker {\em et~al.}(2007)Bussemaker, Foat, and
  Ward]{bussemaker2007}
Bussemaker, H.~J., Foat, B.~C., and Ward, L.~D. (2007).
\newblock Predictive modeling of genome-wide {mRNA} expression: from modules to
  molecules.
\newblock {\em Annu Rev Biophys Biomol Struct\/}, {\bf 36}, 329--347.

\bibitem[Chang {\em et~al.}(2011)Chang, Chang, and Chen]{chang20113820}
Chang, C., Chang, G., and Chen, H. (2011).
\newblock A novel cyclic {AMP/Epac1/CaMKI} signaling cascade promotes {GCM1}
  desumoylation and placental cell fusion.
\newblock {\em Molecular and Cellular Biology\/}, {\bf 31}(18), 3820 -- 3831.

\bibitem[Das {\em et~al.}(2011)Das, Jena, and Levasseur]{Das09122011}
Das, S., Jena, S., and Levasseur, D.~N. (2011).
\newblock Alternative splicing produces {Nanog} protein variants with different
  capacities for self-renewal and pluripotency in embryonic stem cells.
\newblock {\em Journal of Biological Chemistry\/}, {\bf 286}(49), 42690--42703.

\bibitem[Faith {\em et~al.}(2007)Faith, Hayete, Thaden, Mogno, Wierzbowski,
  Cottarel, Kasif, Collins, and Gardner]{FaithJeremiahj}
Faith, J.~J., Hayete, B., Thaden, J.~T., Mogno, I., Wierzbowski, J., Cottarel,
  G., Kasif, S., Collins, J.~J., and Gardner, T.~S. (2007).
\newblock Large-scale mapping and validation of \textit{Escherichia coli}
  transcriptional regulation from a compendium of expression profiles.
\newblock {\em PLoS Biology\/}, {\bf 5}(1), 54--66.

\bibitem[Friedman(2004)Friedman]{NirFriedman02062004}
Friedman, N. (2004).
\newblock Inferring cellular networks using probabilistic graphical models.
\newblock {\em Science\/}, {\bf 303}(5659), 799--805.

\bibitem[Gama-Castro {\em et~al.}(2008)Gama-Castro, Jimenez-Jacinto,
  Peralta-Gil, Santos-Zavaleta, Penaloza-Spinola, Contreras-Moreira,
  Segura-Salazar, Muniz-Rascado, Martinez-Flores, Salgado, Bonavides-Martinez,
  Abreu-Goodger, Rodriguez-Penagos, Miranda-Rios, Morett, Merino, Huerta,
  Trevino-Quintanilla, and Collado-Vides]{SocorroGama-Castro01112008}
Gama-Castro, S., Jimenez-Jacinto, V., Peralta-Gil, M., Santos-Zavaleta, A.,
  Penaloza-Spinola, M.~I., Contreras-Moreira, B., Segura-Salazar, J.,
  Muniz-Rascado, L., Martinez-Flores, I., Salgado, H., Bonavides-Martinez, C.,
  Abreu-Goodger, C., Rodriguez-Penagos, C., Miranda-Rios, J., Morett, E.,
  Merino, E., Huerta, A.~M., Trevino-Quintanilla, L., and Collado-Vides, J.
  (2008).
\newblock {RegulonDB (version 6.0)}: gene regulation model of
  \textit{Escherichia coli} {K-12 beyond transcription, active (experimental)
  annotated promoters and Textpresso navigation}.
\newblock {\em Nucleic Acids Research\/}, {\bf 36}(Database issue), D120--124.

\bibitem[Gasch {\em et~al.}(2000)Gasch, Spellman, Kao, Carmel-Harel, Eisen,
  Storz, Botstein, and Brown]{AudreyP.Gasch12012000}
Gasch, A.~P., Spellman, P.~T., Kao, C.~M., Carmel-Harel, O., Eisen, M.~B.,
  Storz, G., Botstein, D., and Brown, P.~O. (2000).
\newblock Genomic expression programs in the response of yeast cells to
  environmental changes.
\newblock {\em Molecular and Cellular Biology\/}, {\bf 11}(12), 4241--4257.

\bibitem[Harbison {\em et~al.}(2004)Harbison, Gordon, Lee, Rinaldi, Macisaac,
  Danford, Hannett, Tagne, Reynolds, Yoo, Jennings, Zeitlinger, Pokholok,
  Kellis, Rolfe, Takusagawa, Lander, Gifford, Fraenkel, and
  Young]{harbison2004}
Harbison, C.~T., Gordon, D.~B., Lee, T.~I., Rinaldi, N.~J., Macisaac, K.~D.,
  Danford, T.~W., Hannett, N.~M., Tagne, J.~B., Reynolds, D.~B., Yoo, J.,
  Jennings, E.~G., Zeitlinger, J., Pokholok, D.~K., Kellis, M., Rolfe, P.~A.,
  Takusagawa, K.~T., Lander, E.~S., Gifford, D.~K., Fraenkel, E., and Young,
  R.~A. (2004).
\newblock Transcriptional regulatory code of a eukaryotic genome.
\newblock {\em Nature\/}, {\bf 431}, 99--104.

\bibitem[Huynh-Thu {\em et~al.}(2010)Huynh-Thu, Irrthum, Wehenkel, and
  Geurts]{10.1371/journal.pone.0012776}
Huynh-Thu, V.~A., Irrthum, A., Wehenkel, L., and Geurts, P. (2010).
\newblock Inferring regulatory networks from expression data using tree-based
  methods.
\newblock {\em PLoS ONE\/}, {\bf 5}(9), e12776.

\bibitem[Joshi {\em et~al.}(2009)Joshi, De~Smet, Marchal, Van~de Peer, and
  Michoel]{AnaghaJoshi02152009}
Joshi, A., De~Smet, R., Marchal, K., Van~de Peer, Y., and Michoel, T. (2009).
\newblock {Module networks revisited: computational assessment and
  prioritization of model predictions}.
\newblock {\em Bioinformatics\/}, {\bf 25}(4), 490--496.

\bibitem[Karafin {\em et~al.}(2011)Karafin, Parwani, Netto, Illei, Epstein,
  Ladanyi, and Argani]{Karafin20111264}
Karafin, M., Parwani, A.~V., Netto, G.~J., Illei, P.~B., Epstein, J.~I.,
  Ladanyi, M., and Argani, P. (2011).
\newblock Diffuse expression of {PAX2} and {PAX8} in the cystic epithelium of
  mixed epithelial stromal tumor, angiomyolipoma with epithelial cysts, and
  primary renal synovial sarcoma: evidence supporting renal tubular
  differentiation.
\newblock {\em American Journal of Surgical Pathology\/}, {\bf 35}(9),
  1264--1273.

\bibitem[Libório {\em et~al.}(2011)Libório, Acquafreda, Matizonkas-Antonio,
  Silva-Valenzuela, Ferraz, and Nunes]{Tatiana2011225}
Libório, T.~N., Acquafreda, T., Matizonkas-Antonio, L.~F., Silva-Valenzuela,
  M.~G., Ferraz, A.~R., and Nunes, F.~D. (2011).
\newblock In situ hybridization detection of homeobox genes reveals distinct
  expression patterns in oral squamous cell carcinomas.
\newblock {\em Histopathology\/}, {\bf 58}(2), 225--233.

\bibitem[Lukk {\em et~al.}(2010)Lukk, Kapushesky, Nikkila, Parkinson,
  Goncalves, Huber, Ukkonen, and Brazma]{LukkMargus}
Lukk, M., Kapushesky, M., Nikkila, J., Parkinson, H., Goncalves, A., Huber, W.,
  Ukkonen, E., and Brazma, A. (2010).
\newblock A global map of human gene expression.
\newblock {\em Nat Biotech\/}, {\bf 28}(4), 322--324.

\bibitem[Luscombe {\em et~al.}(2004)Luscombe, {Madan Babu}, Yu, Snyder,
  Teichmann, and Gerstein]{luscombe2004}
Luscombe, N.~M., {Madan Babu}, M., Yu, H., Snyder, M., Teichmann, S.~A., and
  Gerstein, M. (2004).
\newblock Genomic analysis of regulatory network dynamics reveals large
  topological changes.
\newblock {\em Nature\/}, {\bf 431}, 308--312.

\bibitem[Marbach {\em et~al.}(2010)Marbach, Prill, Schaffter, Mattiussi,
  Floreano, and Stolovitzky]{Marbach06042010}
Marbach, D., Prill, R.~J., Schaffter, T., Mattiussi, C., Floreano, D., and
  Stolovitzky, G. (2010).
\newblock {Revealing strengths and weaknesses of methods for gene network
  inference}.
\newblock {\em Proceedings of the National Academy of Sciences\/}, {\bf
  107}(14), 6286--6291.

\bibitem[Margolin {\em et~al.}(2006)Margolin, Nemenman, Basso, Wiggins,
  Stolovitzky, Favera, and Califano]{16723010}
Margolin, A., Nemenman, I., Basso, K., Wiggins, C., Stolovitzky, G., Favera,
  R., and Califano, A. (2006).
\newblock {ARACNE}: an algorithm for the reconstruction of gene regulatory
  networks in a mammalian cellular context.
\newblock {\em BMC Bioinformatics\/}, {\bf 7}(Suppl 1), S7.

\bibitem[Michoel {\em et~al.}(2009)Michoel, De~Smet, Joshi, Van~de Peer, and
  Marchal]{Michoel19422680}
Michoel, T., De~Smet, R., Joshi, A., Van~de Peer, Y., and Marchal, K. (2009).
\newblock Comparative analysis of module-based versus direct methods for
  reverse-engineering transcriptional regulatory networks.
\newblock {\em BMC Systems Biology\/}, {\bf 3}(1), 49.

\bibitem[Monteiro {\em et~al.}(2008)Monteiro, Mendes, Teixeira, d'Orey,
  Tenreiro, Mira, Pais, Francisco, Carvalho, Lourenco, Sa-Correia, Oliveira,
  and Freitas]{PedroT.Monteiro01112008}
Monteiro, P.~T., Mendes, N.~D., Teixeira, M.~C., d'Orey, S., Tenreiro, S.,
  Mira, N.~P., Pais, H., Francisco, A.~P., Carvalho, A.~M., Lourenco, A.~B.,
  Sa-Correia, I., Oliveira, A.~L., and Freitas, A.~T. (2008).
\newblock {YEASTRACT-DISCOVERER}: new tools to improve the analysis of
  transcriptional regulatory associations in \textit{Saccharomyces cerevisiae}.
\newblock {\em Nucleic Acids Research\/}, {\bf 36}(suppl\_1), D132--136.

\bibitem[NCBI(????)NCBI]{ncbi}
NCBI (????).
\newblock \url{http://www.ncbi.nlm.nih.gov/}.
\newblock [Online; accessed 01-March-2012].

\bibitem[Perseu {\em et~al.}(2011)Perseu, Satta, Moi, Demartis, Manunza,
  Sollaino, Barella, Cao, and Galanello]{Perseu20102011}
Perseu, L., Satta, S., Moi, P., Demartis, F.~R., Manunza, L., Sollaino, M.~C.,
  Barella, S., Cao, A., and Galanello, R. (2011).
\newblock {KLF1} gene mutations cause borderline {HbA2}.
\newblock {\em Blood\/}, {\bf 118}(16), 4454--4458.

\bibitem[Prill {\em et~al.}(2010)Prill, Marbach, Saez-Rodriguez, Sorger,
  Alexopoulos, Xue, Clarke, Altan-Bonnet, and Stolovitzky]{DREAM3}
Prill, R.~J., Marbach, D., Saez-Rodriguez, J., Sorger, P.~K., Alexopoulos,
  L.~G., Xue, X., Clarke, N.~D., Altan-Bonnet, G., and Stolovitzky, G. (2010).
\newblock Towards a rigorous assessment of systems biology models: The {DREAM3}
  challenges.
\newblock {\em PLoS ONE\/}, {\bf 5}(2), e9202.

\bibitem[Qi {\em et~al.}(2011)Qi, Michoel, and Butler]{qi2011a}
Qi, J., Michoel, T., and Butler, G. (2011).
\newblock Applying linear models to learn regulation programs in a
  transcription regulatory module network.
\newblock {\em Lecture Notes in Computer Science\/}, {\bf 6623/2011}, 37--47.

\bibitem[Schaefer {\em et~al.}(2010)Schaefer, Schmeier, and
  Bajic]{Schaefer21102010}
Schaefer, U., Schmeier, S., and Bajic, V.~B. (2010).
\newblock {TcoF-DB}: dragon database for human transcription co-factors and
  transcription factor interacting proteins.
\newblock {\em Nucleic Acids Research\/}.

\bibitem[Segal {\em et~al.}(2003)Segal, Shapira, Regev, Pe'er, Botstein,
  Koller, and Friedman]{Segal2003Nature}
Segal, E., Shapira, M., Regev, A., Pe'er, D., Botstein, D., Koller, D., and
  Friedman, N. (2003).
\newblock Module networks: identifying regulatory modules and their
  condition-specific regulators from gene expression data.
\newblock {\em Nature Genetics\/}, {\bf 34}(2), 166--176.

\bibitem[Smyth(2004)Smyth]{Gordon2004}
Smyth, G.~K. (2004).
\newblock Linear models and empirical {Bayes} methods for assessing
  differential expression in microarray experiments.
\newblock {\em Statistical Applications in Genetics and Molecular Biology\/},
  {\bf 3}, Article3.

\bibitem[Smyth(2005)Smyth]{limma}
Smyth, G.~K. (2005).
\newblock Limma: linear models for microarray data.
\newblock In {\em Bioinformatics and Computational Biology Solutions using R
  and Bioconductor\/}, pages 397--420. Springer.

\bibitem[Sotoodehnia {\em et~al.}(2011)Sotoodehnia, Isaacs, and \textit{et
  al.}]{Sotoodehnia20111061}
Sotoodehnia, N., Isaacs, A., and \textit{et al.} (2011).
\newblock Common variants in 22 loci are associated with qrs duration and
  cardiac ventricular conduction.
\newblock {\em Nat Genet\/}, {\bf 42}, 1061--4036.

\bibitem[Wang {\em et~al.}(2011)Wang, Venkatesh, Li, Goetz, Mukherjee, Biswas,
  Zhu, Kaubisch, Wang, Pullman, Whitney, Kuro-o, Roig, Shay, Mohammadi, and
  Mani]{wang20103220}
Wang, H., Venkatesh, M., Li, H., Goetz, R., Mukherjee, S., Biswas, A., Zhu, L.,
  Kaubisch, A., Wang, L., Pullman, J., Whitney, K., Kuro-o, M., Roig, A.~I.,
  Shay, J.~W., Mohammadi, M., and Mani, S. (2011).
\newblock Pregnane x receptor activation induces {FGF19}-dependent tumor
  aggressiveness in humans and mice.
\newblock {\em The Journal of Clinical Investigation\/}, {\bf 121}(8),
  3220--3232.

\bibitem[Wang {\em et~al.}(2010)Wang, Zeng, Li, Liu, Li, Li, Zhao, Wei, Wang,
  Li, Feng, He, and Shi]{wang201013622}
Wang, T., Zeng, Z., Li, T., Liu, J., Li, J., Li, Y., Zhao, Q., Wei, Z., Wang,
  Y., Li, B., Feng, G., He, L., and Shi, Y. (2010).
\newblock Common {SNPs} in {Myelin} transcription factor 1-like {MYT1}:
  Association with major depressive disorder in the chinese han population.
\newblock {\em PLoS ONE\/}, {\bf 5}(10), e13662.

\bibitem[Zhu {\em et~al.}(2004)Zhu, Lum, Lamb, GuhaThakurta, Edwards,
  Thieringer, Berger, Wu, Thompson, Sachs, and Schadt]{zhu2004}
Zhu, J., Lum, P.~Y., Lamb, J., GuhaThakurta, D., Edwards, S.~W., Thieringer,
  R., Berger, J.~P., Wu, M.~S., Thompson, J., Sachs, A.~B., and Schadt, E.~E.
  (2004).
\newblock An integrative genomics approach to the reconstruction of gene
  networks in segregating populations.
\newblock {\em Cytogenet Genome Res\/}, {\bf 105}, 363--374.

\end{thebibliography}

\end{document}